%% file: sample-bibliography.tex
\documentclass[sigconf]{acmart}
\makeatletter
\def\@ACM@checkaffil{
    \if@ACM@instpresent\else
    \ClassWarningNoLine{\@classname}{No institution present for an affiliation}%
    \fi
    \if@ACM@citypresent\else
    \ClassWarningNoLine{\@classname}{No city present for an affiliation}%
    \fi
    \if@ACM@countrypresent\else
        \ClassWarningNoLine{\@classname}{No country present for an affiliation}%
    \fi
}
\makeatother
\renewcommand\footnotetextcopyrightpermission[1]{} 
\usepackage{booktabs} 
\usepackage{algorithm}
\usepackage{algpseudocode}
\usepackage{makecell}
\usepackage{graphicx}
\usepackage{float}
\usepackage{amsmath}
\usepackage{algpseudocode}
\setcopyright{rightsretained}

\acmDOI{10.1145/3477314.3507142}

\acmISBN{978-1-4503-8025-6/22/02}

\acmArticle{4}
\acmPrice{15.00}

\begin{document}
\title{cuPSO: GPU Parallelization for Particle Swarm Optimization Algorithms}

\author{Chuan-Chi Wang}
\affiliation{%
  \institution{National Taiwan University}
  \city{Taipei}
  \state{Taiwan}
}
\email{d10922012@ntu.edu.tw}

\author{Chun-Yen Ho}
\affiliation{%
  \institution{National Taiwan University}
  \city{Taipei}
  \state{Taiwan}
}
\email{chunyenher@gmail.com}

\author{Chia-Heng Tu}
 \orcid{0000-0001-8967-1385}
\affiliation{%
  \institution{National Cheng Kung University}
  \city{Tainan}
  \state{Taiwan}
}
\email{chiaheng@ncku.edu.tw}

\author{Shih-Hao Hung}
 \orcid{0000-0003-2043-2663}
\affiliation{%
  \institution{National Taiwan University}
  \city{Taipei}
  \state{Taiwan}
}
\email{hungsh@csie.ntu.edu.tw}

\begin{abstract}
Particle Swarm Optimization (PSO) is a stochastic technique for solving the optimization problem. Attempts have been made to shorten the computation times of PSO based algorithms with massive threads on GPUs (graphic processing units), where thread groups are formed to calculate the information of particles and the computed outputs for the particles are aggregated and analyzed to find the best solution. In particular, the reduction-based method is considered as a common approach to handle the data aggregation and analysis for the calculated particle information. Nevertheless, based on our analysis, the reduction-based method would suffer from excessive memory accesses and thread synchronization overheads. 
In this paper, we propose a novel algorithm to alleviate the above overheads with the atomic functions. 
The threads within a thread group update the calculated results atomically to the intra-group data queue conditionally, which prevents the frequent accesses to the memory as done by the parallel reduction operations. 
Furthermore, we develop an enhanced version of the algorithm to alleviate the synchronization barrier among the thread groups, which is achieved by allowing the thread groups to run asynchronously and updating to the global, lock-protected variables occasionally if necessary. 
Our experimental results show that our proposed algorithm running on the Nvidia GPU is about 200 times faster than the serial version executed by the Intel Xeon CPU. 
Moreover, the novel algorithm outperforms the state-of-the-art method (the parallel reduction approach) by a factor of 2.2. 
\end{abstract}

\maketitle
\pagestyle{plain}

\makeatletter
\newenvironment{megaalgorithm}[1][htb]{%
    \renewcommand{\ALG@name}{Data Structure}
    \renewcommand{\fnum@algorithm}{\fname@algorithm}
   \begin{algorithm}[#1]%
  }{\end{algorithm}}
\makeatother

\input{samplebody-conf}

\bibliographystyle{ACM-Reference-Format}
\bibliography{sample-bibliography}

\end{document}

%% file: samplebody-conf.tex
\section{Introduction}
Particle Swarm Optimization (PSO), introduced in 1995 by Kennedy and Eberhart~\cite{PSO_main}, is a powerful optimization algorithm based on a stochastic optimization technique.
A PSO algorithm searches the optimum of the target function, called \emph{fitness function}, by simulating social behaviors with the particles, e.g., the movement for a bird flock or fish school to look for food. 
Thanks to the versatility of solving various practical problems~\cite{PSO_app1, PSO_app2}, PSO has been evolved into sophisticated variants to improve the optimization performance~\cite{PSO_variant1, PSO_variant2, PSO_variant3} and to accelerate the execution efficiency with different types of hardware platforms, including CPUs~\cite{CPU1}, Field Programmable Gate Arrays (FPGAs)~\cite{FPGA1, FPGA2}, and GPUs~\cite{GPU1}. 

As a population-based metaheuristic algorithm, it is usually implemented by performing the stochastic search process iteratively, which would demand a large number of particle status updates and the target function evaluations when handling a sophisticated problem. This computation hungry nature poses a challenge for real-time applications to dynamic environments. For example, PSO could be used to track moving objects and to determine the human postures in the computer vision applications. Hence, the capability of fast convergence of PSO is critical to fit the real-time requirements. 

To boost the execution performance of PSO algorithms, GPUs have been adopted to solve the PSO problems with a large number of parallel threads. In particular, particles are partitioned into groups to be handled by the underlying thread groups for the status updates and fitness evaluations, and the data calculated by different thread groups are aggregated and analyzed to obtain the best solution, which is the optimization goal for the next iteration. The iterative process stops when achieving the target precision or the specified iteration number. To improve the execution efficiency, the \emph{parallel reductions} are usually adopted~\cite{reduction}, to expedite the data aggregation and analysis process. Nevertheless, the reduction-based methods require a considerable amount of memory accesses to keep the intermediate results and a synchronization operation for waiting the thread groups to write their intermediate results to the memory before the best data is obtained for this iteration. The above two overheads hinder the potential of performance acceleration for solving the PSO problems.



In this paper, we propose a novel algorithm to alleviate the excessive memory access overhead. It is referred to as the \emph{queue} algorithm since the atomic operation is used to protect the conditional accesses to the memory locations when parallel threads within a group update their results, where the behavior of the sequential accesses is similar to that of the queue data structure. In addition, an enhanced algorithm is developed to further improve the execution efficiency by removing the synchronization barrier when the thread groups update their results, which is done by introducing an atomic lock to control the accesses when updating the best data for the current iteration. The enhanced algorithm, referred to as the \emph{queue lock} algorithm, performs especially well for the PSO problem with lower complexity, i.e., the problem domain is formulated as one dimensional space. The proposed algorithms have been implemented and they are tested with the two problem configurations with low and high complexities (i.e., 1 dimension and 120 dimensions) on the machine with the Intel Xeon processor and the Nvidia GTX 1080 Ti GPU. Our results show that it achieves over 200x speedups by using the GPU, compared with the serial version running on the CPU. Besides, the proposed algorithm is faster than the state-of-the-art method (the reduction-based approach~\cite{GPU_reduction}) by a factor of 2.2, in terms of the computation time of the PSO algorithm. The source code of this work is online available\footnote{https://github.com/wang2346581/cuPSO}, and we believe that our efforts further move a step forward for the real-time processing of PSO problems. The contributions of this work are summarized as follows. 
\begin{enumerate}
\item Propose the queue algorithm to reduce the concurrent memory access delays. 
\item Propose the queue lock algorithm to further remove the synchronization overhead. 
\item Implement the above algorithms and compare their results against those achieved by the state-of-the-art methods, i.e., the parallel reduction and the loop-unrolling methods. The experimental results demonstrate the effectiveness and efficiency of our proposed algorithms. 
\end{enumerate}

The rest of this paper is organized as follows. Section 2 describes the related work. Section 3 provides the background of the PSO algorithms for serial and parallel executions. Section 4 presents the design of our proposed algorithms. Section 5 details the implementation remarks. Section 6 shows the experimental results. Section 7 concludes this paper and gives the future work. 

\section{Particle Swarm Optimization (PSO) Algorithms}
\subsection{Standard PSO Algorithm (SPSO)} \label{sec:spso}
The parameters of the Standard Particle Swarm Optimization (SPSO) algorithm are defined in Table~\ref{tab:parameter_definition}. In general, the parameters $w$, $c1$, and $c2$ are changed in different practical situations, controlling the behavior and efficiency of the SPSO algorithm. 
$max\_pos$, $min\_pos$, $max\_v$, and $min\_v$ restrict the lower and upper boundaries of each dimension in the search space for the position and velocity respectively.
$max\_iter$ performs the termination criterion of the number of iterations. 
$particle\_cnt$ is the total number of the particles.
Moreover, $pos_i$, $v_i$, $fit_i$, $pbest\_pos_i$, and $pbest\_fit_i$ are the necessary information of each particle$_i$, which represent as the position, speed, fitness, best-known position, and best-known fitness respectively. The subscript $i$ is the index of each particle.
Finally, $gbest\_pos$ is the best-fitness point ever found by the whole swarm, and $gbest\_fit$ is the best-fitness value obtained by corresponding best-fitness point $gbest\_pos$.

\begin{table}
  \caption{Description of PSO required parameters.}
  \label{tab:parameter_definition}
  \begin{tabular}{p{1.5cm}p{6.2cm}}
    \toprule
    Name & Description \\
    \midrule
    $w$ & the inertia weight \\
    $c1$ & the cognitive coefficient \\
    $c2$ & the social coefficient \\
    $max\_pos$ & the maximum position of the 
    particles \\
    $min\_pos$ & the minimum position of the 
    particles \\
    $max\_v$ & the maximum velocity of the particles \\
    $min\_v$ & the maximum velocity of the particles \\
    $max\_iter$ & the number of the iterations \\
    $particle\_cnt$ & the total number of the particles \\
    $pos_i$ & the position of each particle$_i$ \\ 
    $v_i$ & the velocity of each particle$_i$ \\ 
    $fit_i$ & the fitness of each particle$_i$ \\
    $pbest\_pos_i$ & the best-known position of each particle$_i$ \\
    $pbest\_fit_i$ & the best-known fitness of each particle$_i$ \\
    $gbest\_pos$ & the best-known position of the whole swarm \\
    $gbest\_fit$ & the best-known fitness of the whole swarm \\
\bottomrule
\end{tabular}
\end{table}

Algorithm~\ref{algo:spso_cpu} shows the full workflow of the SPSO algorithm, and it could be simply decomposed into 5 steps. 
The algorithm would continue operating from step 2 to step 5 until satisfying the termination condition as follows:
(1) initialize the information of each particle, including position, velocity, fitness, the best-known position of each particle, the best-known fitness of each particle, global best-known position, and global best-known fitness. 
The position and velocity are initialized with a uniformly distributed random vector, constricted in the fixed range to ensure the convergence. 
(2) update the velocity of each particle as Equation~\ref{eq:update_v} and the position of each particle as Equation~\ref{eq:update_pos} within the domain of the fitness function at time $t$ to explore the whole search-space.
r1 and r2 are uniformly distributed on random value in [0, 1]. 
(3) calculate each particle fitness by its position and fitness function. 
(4) update the best-known position and the fitness, local best for short, while the new fitness is better than the previous one. 
(5) update the global best-known position and fitness, global best for short, while the fitness of the particle is better than the global best-known fitness.

\begin{equation}
\begin{split}
\label{eq:update_v}
v_i(t+1) = &w*v(t) + c1*r1(pbest\_pos_i(t) - pos_i(t)) \\ &+ c2*r2(gbest\_pos_i(t)-pos_i(t))
\end{split}
\end{equation}

\begin{equation}
\label{eq:update_pos}
pos_i(t+1) = pos_i(t) + v_i(t+1)
\end{equation}

\begin{algorithm}
\caption{Sequential SPSO algorithm}
  \label{algo:spso_cpu}
  \begin{algorithmic}[1]
    \For{$i$ in $particle\_cnt$}
    \Comment{Step1: Initialization}
        \State Initialize $pos_i$ and $v_i$ randomly.
        \State Evaluate  $fit_i$ by $pos_i$ and fitnessfunction.
        \State Initialize $pbest\_fit_i$ and $pbest\_pos_i$.
        \State Update $gbest\_pos$ and $gbest\_fit$.
    \EndFor
    \For{$j$ in $max\_iter$}
    \Comment{Step2-5: Computation}
        \For{$i$ in $particle\_cnt$}
            \State Update $v_i$ by $pos_i$, $pbest\_pos_i$, and $gbest\_pos$. \Comment{Step2}
            \State Keep $v_i$ in the range $(min\_v_i, max\_v_i)$
            \State Update $pos_i$ by $pos_i$ and $v_i$.
            \State Keep $pos_i$ in the range $(min\_pos_i, max\_pos_i)$.
            \State Evaluate $fit_i$ by $pos_i$. \Comment{Step3}
            \If{$fit_i \textgreater pbest\_fit_i$} \Comment{Update local best}
                \State Update $pbest\_fit_i$ and $pbest\_pos_i$ by $fit_i$ and $pos_i$. 
            \EndIf \Comment{Step4}
            \If{$pbest\_fit_i \textgreater gbest\_fit$} \Comment{Update global best}
                \State Update $gbest\_pos$ and $gbest\_fit$ by $pbest\_pos_i$ and $pbest\_fit_i$ 
            \EndIf \Comment{Step5}
             
        \EndFor
    \EndFor
  \end{algorithmic}
\end{algorithm}
\subsection{Parallel PSO Algorithm (PPSO)} \label{sec:ppso}
In order to take advantage of the computing power of GPUs (graphics processing units), a parallel PSO (PPSO) has been developed based on the SPSO introduced in Section~\ref{sec:spso} to improve the execution speed of Steps 2-5 in Algorithm~\ref{algo:spso_cpu}. The PPSO program is executed by CPU and GPU. Specifically, the particle information initialization (Step 1) and finalization (to obtain the final results) are done by the CPU, whereas the compute-intensive part (Steps 2-5) is performed on the GPU.

\begin{figure}[htb!]
\centering
\includegraphics[width=0.45\textwidth]{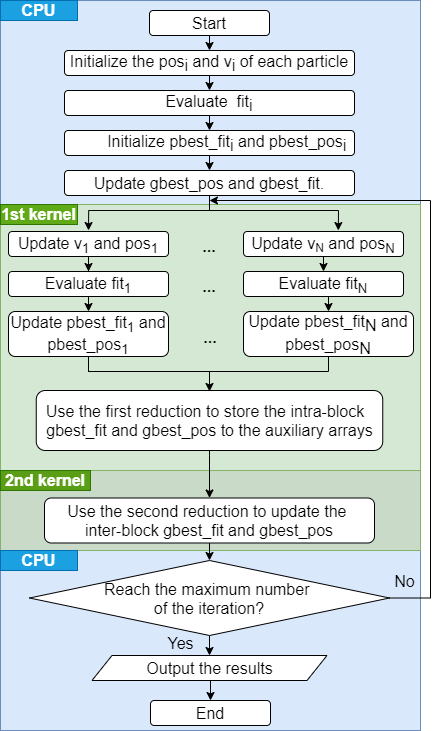} 
\caption{The workflow of parallel PSO algorithm.}
\label{fig:algo_flow}
\end{figure}

In PPSO~\cite{GPU_reduction}, a thread is used to handle the computations for calculating the information of a particle, e.g., velocity, position, and fitness, as defined in Steps 2 and 3, where the workflow is illustrated in the "1st kernel" (the first parallel code section) in Figure~\ref{fig:algo_flow}. 
The threads are grouped into \emph{blocks} to update the best-known position and fitness for a group of particles, where the best data within each block is computed at the end of the "1st kernel" in Figure~\ref{fig:algo_flow}. It is important to note that in order to speed up the process of obtaining the best data within each block, the \emph{parallel reduction} operations are performed within each block (i.e., each thread group). Subsequently, the second parallel code (the "2nd kernel") is used to further derive the best position and fitness for the particle data across the thread groups (blocks), which is also performed by the parallel reductions.
It is worthy to note that the first kernel code is performed by multiple thread blocks and each block performs the reductions internally to find the \emph{local} best data points. 
On the contrary, the second kernel code is handed by a thread block to search for the \emph{global} best data point from the output data derived by the thread blocks for the first kernel. 

During the parallel reduction, it incurs extensive arithmetic operations for calculating the proper addresses of the reduction operands and the branching instructions in the reduction loops. 
A common approach is to unroll the reduction loops to reduce the above overheads since the addresses are computed offline and the sequential unrolled code avoids the branches.
Nevertheless, unrolling the loops cannot remove the implicit synchronization between the first and second kernels, and during the iterative search process, the two kernels are invoked repeatedly, which introduces a significant amount of overheads. In this work, we develop the novel algorithm to tackle the problem.


\section{Proposed Parallel Algorithms}
\subsection{PPSO with the Shared Memory Queue}
\label{sec:queue}
In the practical test, we find that it is seldom the case that the conditions of Steps 4 and 5 in Algorithm~\ref{algo:spso_cpu} are satisfied; the possibility of the satisfied condition may be less than 0.1\%.
Specifically taking Step 4 as an example, the $fit_{i}$ calculated by the $i$-th thread is usually less than the $pbest\_fit_{i}$ listed in line 14 of the algorithm. Based on the observation, we develop the shared memory based solution, which is referred to as the \emph{queue} algorithm, to let each thread to update the queue (the in-order accessed arrays and the index to the arrays guarded by the atomic addition operation) if its calculated $fit$ value is larger than that of the global value $gbest\_fit$. The major concept is the sequential accesses (enforced by the atomic add operation) to the shared memory arrays and is listed in line 1-4 in Algorithm~\ref{algo:spso_queue}. 
To further identify the best configuration of each block, the thread with its thread identifier equal to zero (i.e., $threadIdx.x == 0$) is selected to scan the entire queue, finds the local best configuration, and updates the local best information to the auxiliary arrays, as listed in line 7-20 in Algorithm~\ref{algo:spso_queue}. In particular, the local best information for a block (stored in $bestFitQueue[0]$) is updated to the auxiliary arrays indexed by the block identifier ($auxFit[blockIdx.x]$ and $auxPos[blockIdx.x]$), as specified in line 17-18  Algorithm~\ref{algo:spso_queue}.
Regarding the complexity of the step for updating the intra-block best, we reduce the complexity from $\mathcal{O}(\log{} n)$ of using the reduction operation to theoretical $\mathcal{O}(1)$ of using the queue method\footnote{Thanks to the queue method that maintains the best-known fitness of each particle and performs the reduction operation if necessary, the resultant complexity is close to $\mathcal{O}(1)$ at most of the time.}. 
Note that the similar approach is also adopted when using the second kernel to update the global best information.

\begin{algorithm}[htb!]
\caption{The queue algorithm.}
  \label{algo:spso_queue}
  \begin{algorithmic}[1]
  \If{$fit > gbest\_fit$}
    \State $unsigned\;const\;qIdx \gets atomicAdd(
    \&num, 1);$
    \State $bestFitQueue[qIdx] \gets fit;$
    \State $bestPosQueue[qIdx] \gets pos;$
  \EndIf
  \State \_\_syncthreads();
  \If{$idx < particle\_cnt\;\&\&\;threadIdx.x==0$}
    \State $auxFit[blockIdx.x] \gets INT\_MIN;$
    \State $auxPos[blockIdx.x] \gets INT\_MIN;$
    \If{$num$}
      \For{$j$ in $num$}
        \If{$bestFitQueue[j] > bestFitQueue[0]$}
          \State $bestFitQueue[0] \gets bestFitQueue[j]$
          \State $bestPosQueue[0] \gets bestPosQueue[j]$
        \EndIf
      \EndFor
      \State $auxFit[blockIdx.x] \gets bestFitQueue[0];$
      \State $auxPos[blockIdx.x] \gets bestPosQueue[0];$
    \EndIf
  \EndIf
  \end{algorithmic}
\end{algorithm}

\subsection{PPSO with the Shared Memory Queue and the Global Lock} \label{sec:queue-lock}
As described in Section \ref{sec:queue}, the local best information of each block is stored in the corresponding position of the auxiliary arrays (line 17-18 of Algorithm~\ref{algo:spso_queue}). This design allows the first threads of each block to store their best data to the auxiliary arrays concurrently. Later, the second kernel is used to further search for the global best information from within the auxiliary arrays. 
In such a design, it requires the two kernels to compute the global best configuration, as illustrated in Figure \ref{fig:algo_flow}. 
In order to alleviate the synchronization overhead (between the two kernels) and shorten the time of launching the second kernel,
we fuse the two kernels together via computing the best configuration (originally done in the second kernel) at the end of the first kernel. That is, Algorithm~\ref{algo:spso_queueLock} is proposed to replace the line 17-18 of Algorithm~\ref{algo:spso_queue}.

Algorithm~\ref{algo:spso_queueLock} reduces the 
execution
time because each thread block compares its local best configuration against the global best one as soon as its local best configuration is available. 
In the fused kernel, each of the thread blocks finds its local best configuration and updates to the global configuration if necessary. 
Such a design eliminates the need to write to the auxiliary arrays and relaxes the synchronization of the thread blocks (i.e., the second kernel should be started after the threads in the first kernel are terminated), where a lock is introduced to ensure the global best information ($gbest\_fit$ and $gbest\_pos$) is accessed sequentially by utilizing the atomic lock. 
Therefore, it would deliver better performance, especially for the one dimensional 
data for $gbest\_fit$, which has no bad side effort.

\begin{algorithm}[hbt!]
\caption{The queue algorithm with atomic lock.}
  \label{algo:spso_queueLock}
  \begin{algorithmic}[1]
  \State while(atomicCAS(lock, 0, 1) != 0);
  \If{$bestFitQueue[0] > gbest\_fit$}
    \State $gbest\_fit \gets bestFitQueue[0];$
    \State $gbest\_pos \gets bestPosQueue[0];$
    \State \_\_threadfence();
  \EndIf
  \State atomicExch(lock, 0);
  \end{algorithmic}
\end{algorithm}

\section{Implementation Remarks}
\subsection{Coalesce Memory Access}
In general, the developers often use an array of structures (AoS) layout to design a program, since the AoS layout is easily understood and supported directly by most programming languages.
However, AoS layout is almost the worst case scenarios to develop a parallel program as CUDA.
On the other hand, structure of arrays (SoA) layout can reduce memory access time because each thread can do the coalesced access for continuous memory address.
Accordingly, we change the data structure as the SoA layout in our implementation.

\begin{figure}
\centering
\includegraphics[width=0.5\textwidth]{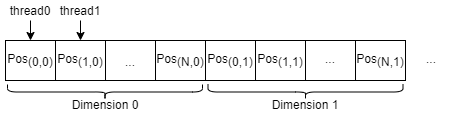} 
\caption{The SoA layout of particle position in high dimension with coalesce memory access}
\label{fig:SoA_high}
\end{figure}

The data structure how we implement the SoA layout of the PSO algorithm.
With the SoA adjustment, every element is seen as to be saved in an array with a fixed value, and it guarantees all threads of the same warp that could access consecutive values in global memory and obtain a better performance than AoS layout.
In the high dimension case of the PSO algorithm, we follow the SoA rule in our implementation. 
As Figure~\ref{fig:SoA_high}, all threads accessing at the same dimension, which are still satisfied with the condition of the coalesced memory access, making the memory access each element efficiently.

\begin{megaalgorithm}
\caption{AoS: Array of structures.}
  \label{algo:array_of_structs}
  \begin{algorithmic}[1]
  \State struct Particle$\lbrace$
  \State $\qquad$ double position;
  \State $\qquad$ double velocity;
  \State $\qquad$ double fitness;
  \State $\qquad$ double pbest\_pos;
  \Comment{The local best-known position}
  \State $\qquad$ double pbest\_fit;
  \Comment{The local best-known fitness}
  \State $\rbrace $ particles[N];
  \end{algorithmic}
\end{megaalgorithm}

\begin{megaalgorithm}
\caption{SoA: Structure of arrays.}
  \label{algo:struct_of_arrays}
  \begin{algorithmic}[1]
  \State struct ParticleCoalescedMem$\lbrace$
  \State $\qquad$ double position[N];
  \State $\qquad$ double velocity[N];
  \State $\qquad$ double fitness[N];
  \State $\qquad$ double pbest\_pos[N];
  \Comment{The local best-known position}
  \State $\qquad$ double pbest\_fit[N];
  \Comment{The local best-known fitness}
  \State $\rbrace$;
  \end{algorithmic}
\end{megaalgorithm}

\subsection{Constant Memory}
NVIDIA GPUs provide 64KB of constant memory that is treated differently from standard global memory.
The constant memory has its own cache. 
When all threads access the same memory address simultaneously and threads in the same block re-access that address continuously, the constant memory can reach the most effective performance. 
Thus, using constant memory instead of global memory, which can conserve the memory bandwidth.
Although the constant memory can complete the memory accessing task with the less latency, it constrains the data usage to be read-only.
In our study case, we can place the parameters $w$, $c1$, $c2$, $max\_pos$, $min\_pos$, $max\_v$, $min\_v$, $particle\_cnt$ in constant memory, making all threads accessed to the constant memory in the same time, with an eye to reducing the memory latency time.

\subsection{Shared Memory}
The shared memory buffers reside physically on the GPU as opposed to residing in off-chip DRAM, which is much faster than the global memory.
In fact, if there are no bank conflicts, the shared memory latency is about 100 times lower than the uncached global memory latency.
Because the shared memory is shared by threads in a thread block, it has a mechanism for threads to cooperate.
As subsection~\ref{sec:ppso}, using reduction would access the global memory frequently so that we can utilize shared memory to reduce the memory bandwidth latency.
Similarly, shared memory can also be used on the GPU Queue method proposed at subsection~\ref{sec:queue}.
Owing to shared memory providing 48KB for each stream multiprocessors (SMs) only, we should store the index of the particle number rather than all axes positions on high dimension cases.
After getting the best fitness of the particle, we can use the index to fill all axes position back, which not only saves the latency time but also reduces the shared memory usage to ensure maximum parallelization for SMs.

\subsection{Random Number Generation}
The PSO algorithm relies on the sequence of random numbers to find the best solution within the designated space. To generate the high-quality pseudorandom and quasirandom numbers on the GPU side, it is important to use the cuRAND~\cite{cuRAND} library offered by the CUDA toolkit. While the users can provide the random number generator function on the GPU side by porting that function seen in the CPU side, it is time-consuming and inefficient for GPU execution since the custom-made library should be a thread-safe implementation and it may not take advantage from the low-level primitives offered by the CUDA toolkit. Based on our experiments, the cuRAND library outperforms the custom-made implementation on the PPSO program by a factor of 1.1. Therefore,in our experimental results, we adopt the cuRAND library (i.e., the $curand\_uniform\_double()$ function) to run the experiments by default.  
%

\section{Experimental Results}
\subsection{Experimental Setup}
The hardware and software platforms used for the experiments are listed in Table \ref{tab:platform}. The five PSO algorithms have been implemented and tested in the experiments, 1) \texttt{CPU} for the serial version defined in Section~\ref{sec:spso}, 2) \texttt{Reduction} for the parallel reduction method introduced in Section~\ref{sec:ppso}, 3) \texttt{Loop Unrolling} for the loop unrolling method introduced in Section~\ref{sec:ppso}, 4) \texttt{Queue} for the proposed queue algorithm defined in Section~\ref{sec:queue}, and 5) \texttt{Queue Lock} for the proposed queue lock algorithm defined in Section~\ref{sec:queue-lock}, where all the algorithms are running on the GPU, except for the first serial version. The \emph{double} precision floating point numbers are adopted in the PSO implementations. Each of the reported data is the average numbers of the execution time for 10 runs, removing the maximum and minimum numbers. In order to observe the impacts of the improved algorithms, the running time at the GPU side is measured and reported. The above implementations are used to solve the problems with both the low and high dimension space, denoted as 1D problem and 120D problem, respectively. 


\begin{table}[hbt!]
  \caption{The hardware and software platforms for the experiments.}
  \label{tab:platform}
  \begin{tabular}{p{1.3cm}p{6.5cm}}
    \toprule
    Name & Description \\
    \midrule
    CPU & Intel Xeon CPU E3-1275 v5, 3.60GHz \\
    GPU & Nvidia GTX-1080ti, 1481 MHz, 3584 CUDA cores, Compute Capability 6.1\\
    OS  & Ubuntu 18.04.4 LTS (kernel version 5.4.0-42-generic)\\
    CUDA & CUDA Toolkit 11.2 \\
    \bottomrule
\end{tabular}
\end{table}

\paragraph{Fitness Function} An important design parameter of the PSO algorithm is the fitness function. While there are several different functions to be used, such as Sphere, Rosenbroc, Griewang, we choose the Cubic function as the fitness function for all the experiments, as shown in Equation~\ref{eq:fit}, since it requires slightly higher computation complexity (e.g., higher than Sphere).  The parameter $w$ used by the fitness function is set as 1 and learning factor $c1$ and $c2$ as 2, which are commonly seen settings.
\begin{equation}
\label{eq:fit}
f = \sum_{i=1}^{d} x_{i}^{3} - 0.8x_{i}^{2} - 1000x_{i} + 8000, \quad -100 \leq x_{i} \leq 100
\end{equation}

\subsection{1D Problem}
\begin{table*}
   \caption{The execution times of the five implementations on the 1D problem.}
  \label{tab:methods}
  \centering
  \begin{tabular}{ccccccc}
    \toprule
    Particles & Iteration & CPU (s) &
    GPU Reduction (s) & 
    GPU Loop Unrolling (s) &
    GPU Queue (s) &
    GPU Queue Lock (s) 
    \\
    \midrule
    32    & 100,000 & 0.100 & 0.413 & 0.394 & 0.368 & {\bf 0.216}  \\
    64    & 100,000 & 0.187 & 0.419 & 0.402 & 0.368 & {\bf 0.219}  \\
    128   & 100,000 & 0.385 & 0.447 & 0.408 & 0.371 & {\bf 0.220}  \\
    256   & 100,000 & 0.825 & 0.455 & 0.419 & 0.371 & {\bf 0.222}  \\
    512   & 100,000 & 1.503 & 0.467 & 0.422 & 0.391 & {\bf 0.223} \\
    1,024 & 100,000 & 3.042 & 0.491 & 0.439 & 0.394 & {\bf 0.227}  \\
    2,048 & 100,000 & 6.277 & 0.508 & 0.451 & 0.409 & {\bf 0.230}  \\
    \bottomrule
  \end{tabular}
\end{table*}

\begin{figure}
\centering
\includegraphics[width=0.5\textwidth]{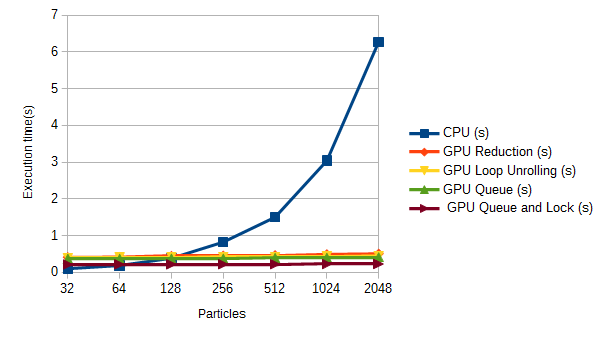}
\caption{Plotting of the execution times of the five implementations with different number of particles involved.}
\label{fig:time}
\end{figure}

As listed in Table~\ref{tab:methods}, the parallel algorithms running on the GPU outperform the serial version on the CPU. The execution time of the CPU version increases almost linearly with the growth of the computation load (related to the number of particles and iteration). On the contrary, the parallel versions have relatively stable execution times across the different workloads since the problem size is relatively small which can be covered by GPU. Specifically, each particle is assigned with a GPU thread to handle the required computation. The maximum number of particles is 2,048 in this experiment, and the Nvidia GPU has 3,584 CUDA threads. The ranking, in terms of the execution time, of the five algorithms are plotted in Figure~\ref{fig:time}. Obviously, the Queue Lock algorithm achieves the best performance than all other parallel versions, and the Queue algorithm is the second best choice. Particularly, the Queue Lock algorithm is 2.2 times faster than the Reduction algorithm, which is the state-of-the-art work. Table~\ref{tab:low_dim_result} shows the speedup ratio of the serial version running on the CPU to the Queue Lock algorithm on the GPU. It shows a maximum of 195x speedup when handling 65,536 particles, and the speedup drops for handling 131,072 particles. Based on our analysis, the GPU is overloaded in such a case, where the GPU capacity is full and it cannot handle the workload efficiently (context switching among thread groups do not help the performance). 

\begin{table}[htb!]
  \caption{The speedups achieved by the Queue Lock algorithm on the 1D problem.}
  \label{tab:low_dim_result}
  \begin{tabular}{cccccc}
    \toprule
    Particles & Iteration & CPU (s) & GPU QueueLock(s) & Speedup Ratio \\
    \midrule
    128     & 100,000 & 0.385 & 0.220 & 1.75 \\
    256     & 100,000 & 0.825 & 0.222 & 3.71 \\
    512     & 100,000 & 1.503 & 0.223 & 6.73 \\
    1,024   & 100,000 & 3.042 & 0.227 & 13.40 \\
    2,048   & 100,000 & 6.277 & 0.230 & 27.29 \\
    4,096   & 100,000 & 12.410 & 0.265 & 46.83 \\
    8,192   & 100,000 & 23.850 & 0.316 & 75.47 \\
    16,384  & 100,000 & 47.355 & 0.417 & 113.56 \\
    32,768  & 100,000 & 94.629 & 0.643 & 147.16 \\
    65,536  & 100,000 & 200.536 & 1.026 & 195.45 \\
    131,072 & 100,000 & 378.671 & 2.759 & 137.24 \\
    \bottomrule
\end{tabular}
\end{table}

\subsection{120D Problem}
We configure the PSO algorithms to handle the problem defined by the 120 dimensional space. In the experiment setup, we select the Queue algorithm to compare against the serial version since the time saved by the Queue Lock algorithm is negligent in the high dimension space since the execution of the first kernel dominate the total execution time and the synchronization overhead saved by the Queue Lock is relatively small. Hence, we think the Queue algorithm is a better solution for the high dimension problem. The speedups achieved by the Queue algorithm are listed in Table \ref{tab:high_dim_result}. Because of a higher dimension space, the peak speedup is achieved (225x) when the number of the particles is 32,768, which is only the half of the number (65,535) reported in Table~\ref{tab:low_dim_result}.


\begin{table}
  \caption{The speedups achieved by the Queue algorithm on the 120D problem.}
  \label{tab:high_dim_result}
  \begin{tabular}{cccccc}
    \toprule
    Particles & Iteration & CPU (s) & GPU Queue(s) & Speedup Ratio \\
    \midrule
    128     & 5,000 & 2.392   & 0.487 & 4.91 \\
    256     & 4,000 & 3.543   & 0.384 & 9.22 \\
    512     & 3,000 & 5.305   & 0.288 & 18.42 \\
    1,024   & 2,000 & 7.078   & 0.225 & 31.45 \\
    2,048   & 2,000 & 14.214  & 0.255 & 55.74 \\
    4,096   & 1,500 & 21.593  & 0.220 & 98.15 \\
    8,192   & 1,000 & 29.494  & 0.191 & 154.41 \\
    16,384  & 1,000 & 59.125  & 0.294 & 201.10 \\
    32,768  & 1,000 & 128.349 & 0.570 & 225.17 \\
    65,536  & 1,000 & 237.933 & 1.169 & 203.53 \\
    131,072 &   800 & 379.820 & 1.744 & 217.78 \\
    \bottomrule
\end{tabular}
\end{table}

\section{Conclusion and Future Work}
In this work, we propose the queue-based algorithms to shorten the execution time for solving the PSO problem. 
We have shown the key ideas of the parallelizing  algorithms for GPUs. Our experimental results show that our proposed algorithms can effectively improve the execution efficiency for solving the PSO problems, achieving 200x speedup compared with the serial version and 2.2x speedup compared with the parallel reduction based method. 
With the encouraging results, in the future, we would like to further improve the performance of the queue-based approach with the asynchronous execution scheme and to extend the algorithm for the multiple GPU version so as to handle a larger size of PSO problems.